\documentclass[prd,nofootinbib,preprint,epsfig]{revtex4}
\usepackage{psfig}
\begin{document}
\def\wng{{{\cal W}}_0^\gamma}
\def\wnz{{{\cal W}}_0^Z}
\def\wcg{{{\cal W}}_c^\gamma}
\def\wcz{{{\cal W}}_c^Z}
\def\wuz{{{\cal W}}_1^Z}
\def\wdz{{{\cal W}}_2^Z}
\def\wtz{{{\cal W}}_3^Z}
\def\zng{{{\cal Z}}_0^\gamma}
\def\zcg{{{\cal Z}}_c^\gamma}
\def\znz{{{\cal Z}}_0^Z}
\def\zcz{{{\cal Z}}_c^Z}
\preprint{\hbox{YITP-SB-54-03}}
\title{Bosonic Quartic Couplings at LHC}

\author{O.\ J.\ P.\ \'Eboli   \footnote{E-mail: eboli@fma.if.usp.br}}
\affiliation{Instituto
de F\'{\i}sica da USP, C.P. 66.318, S\~ao Paulo, SP 05315-970, Brazil.}

\author{M.C. Gonzalez-Garcia
\footnote{E-mail: concha@insti.physics.sunysb.edu}}
\affiliation{ Y.I.T.P., SUNY at Stony Brook, Stony Brook, NY 11794-3840, USA\\
IFIC, Universitat de Val\`encia - C.S.I.C., Apt 22085, 46071
Val\`encia, Spain}

\author{S.\ M.\ Lietti \footnote{E-mail: lietti@fma.if.usp.br}}
\affiliation{Instituto
de F\'{\i}sica da USP, C.P. 66.318, S\~ao Paulo, SP 05389-970, Brazil.}
%\vspace*{1cm}}

%---------------------------------------------------------------------

\begin{abstract}
\vspace*{1cm}

We analyze the potential of the CERN Large Hadron Collider (LHC) to
study anomalous quartic vector--boson interactions $Z Z \gamma
\gamma$, $Z Z Z \gamma$, $W^+ W^- \gamma \gamma$, and $W^+ W^- Z
\gamma$ through the weak boson fusion processes $ q q\to q q \gamma
\gamma$ and $ q q \to q q \gamma Z(\to \ell^+ \ell^-)$ with $\ell = e$
or $\mu$. After a careful study of the backgrounds and how to extract
them from the data, we show that the process $p p \to j j \gamma l^+
l^-$ is potentially the most sensitive to deviations from the Standard
Model, improving the sensitivity to anomalous couplings by up 
to a factor $10^4$ ($10^2$) with respect to  the present direct
(indirect) limits.

\end{abstract}

\maketitle

%---------------------------------------------------------------------
\section{Introduction}

Within the framework of the Standard Model (SM), the structure of the
trilinear and quartic vector--boson couplings is completely determined
by the $SU(2)_L \times U(1)_Y$ gauge symmetry. The study of these
interactions can either lead to an additional confirmation of the
model or give some hint on the existence of new phenomena at a higher
scale~\cite{anomalous}. The triple gauge--boson couplings have been
probed at the Tevatron~\cite{teva} and LEP~\cite{lep,ewwg} through the
production of vector--boson pairs, however, we have only started to
study directly the quartic gauge--boson couplings~
\cite{quarticnow,ewwg,Belanger:1999aw,stir1}. If any deviation from
the SM predictions is observed, independent tests of the triple and
quartic gauge--boson couplings can give important information on the
type of New Physics (NP) responsible for the deviations. For example,
the exchange of heavy bosons can generate a tree level contribution to
four gauge--boson couplings while its effect in the triple--gauge
vertex would only appear at one one--loop, and consequently be
suppressed with respect to the quartic one. Further information on the
NP dynamics can also be provided by determining whether NP reveals
itself in the form of anomalous four-gauge couplings involving only
weak gauge bosons or in those involving photons or in both.

At present the scarce experimental information on quartic anomalous
couplings arises from the processes $e^+e^-\rightarrow W^+W^-\gamma$,
$ Z\gamma\gamma$, $ Z Z \gamma$, and $ \nu \bar{\nu} \gamma \gamma$ at
LEP~\cite{lep,ewwg}.  Due to phase space limitations, the best
sensitivity is attainable for couplings involving photons which should
appear in the final state.  Photonic quartic anomalous couplings can
also affect $\gamma\gamma Z$ and $\gamma\gamma W$ productions at
Tevatron~\cite{our:quartic,stir2}, however, it was shown in
Ref.~\cite{our:quartic} that even with an integrated luminosity of $2$
fb$^{-1}$, the Tevatron experiments can only probe the gauge quartic
couplings at the level of precision obtained at LEP. In the near
future, both photonic and non-photonic quartic gauge couplings will be
tested in pair production of gauge bosons at the LHC via weak boson
fusion (WBF)~\cite{our:quartic,vvvv}.  In the long term, high
sensitivity to anomalous photonic four--gauge couplings is expected at
a next $e^+ e^-$ linear collider ~\cite{Belanger:1999aw,nlc}, as well
as high energy $\gamma\gamma$ ~\cite{bel:bou,ggnos}, and $e\gamma$
~\cite{our:vvv} colliders.

In this work, we study the potential of the LHC to probe the photonic
quartic vertices $Z Z \gamma \gamma$, $W^+ W^- \gamma \gamma$, $W^+
W^- Z \gamma$, and $Z Z Z \gamma$.  The motivation for this study is
two folded: first, even at LHC energies, the best experimental
sensitivity is expected for couplings involving photons due to phase
space limitations. Second, if a signal is observed, the comparison of
the processes here studied, which are only sensitive to photonic
quartic operators, with the observations for processes also dependent
on non-photonic couplings, such as weak gauge boson pair production,
could reveal some symmetries of the underlying dynamics.

We perform a detailed analysis of the most sensitive channels that are
the production via WBF of photon pairs accompanied by jets, {\em
  i.e.},
\begin{equation}
p + p \to q + q \to j + j + \gamma + \gamma   \; ,
\label{jj}
\end{equation}
and the WBF production of a pair of jets plus a photon accompanied by
a lepton pair, where the fermions originate from the decay of a $Z^0$
or a virtual photon {\it i.e.}
\begin{equation} 
p + p \to q  q \to j + j + \gamma + (Z^* \text{ or } \gamma^* \to) \; 
\ell^+ + \ell^-   \; ,
\label{ll}
\end{equation} 
with $\ell = e$ or $\mu$. The advantage of WBF, where the scattered 
final-state quarks receive significant transverse momentum and are 
observed in the detector as far-forward/backward jets, is the 
strong reduction of QCD backgrounds due to the kinematical configuration 
of the colored part of the event.

The process depicted in Eq.~(\ref{ll}) receives contributions from all
four-gauge-boson vertices that we are interested in, while only the $Z
Z \gamma \gamma$ and $W^+ W^- \gamma \gamma$ vertices are relevant for
the process in Eq.~(\ref{jj}). We previously studied the reaction
(\ref{jj}) in Ref.~\cite{our:quartic}. Here, we revisit the limits
there obtained after taking careful account of the QCD uncertainties
in the background evaluation and analyzing strategies to minimize it,
and compare them with the expected sensitivity from (\ref{ll}).

This paper is organized as follows. We present in Sect.~II the
effective operators we analyzed in this work. Section III contains our
analysis of the signal and backgrounds, as well as the attainable
limits ate the LHC. We draw our conclusions in Sect.~VI.

%---------------------------------------------------------------------
\section{Effective quartic interactions}

We parameterize in a model independent form the possible deviations of
the SM predictions for the photonic quartic gauge couplings with the
assumptions that NP respects $SU(2)_L \times U(1)_Y$ gauge invariance
and that no new heavy resonance has been observed.  In this scenario
the gauge symmetry is realized nonlinearly by using the chiral
Lagrangian approach as in Ref.~\cite{Belanger:1999aw}.  Following the
notation of Ref.~\cite{Appelquist}, the building block of the chiral
Lagrangian is the dimensionless unimodular matrix field $\Sigma(x)$,
\begin{equation}
     \Sigma(x) ~=~ \exp\left[ i \frac{\varphi^a(x) \tau^a}{v}\right] \; ,
\end{equation}
where $v=(\sqrt{2}G_F)^{-1}$.  The $\varphi^a$ fields are the would-be
Goldstone fields and $\tau^a$ ($a=1$, $2$, $3$) are the Pauli
matrices.  The $SU(2)_L \times U(1)_Y$ covariant derivative of
$\Sigma$ is defined as
\begin{equation}
D_\mu \Sigma ~\equiv~ \partial_\mu \Sigma 
+ i g \frac{\tau^a}{2} W^a_\mu \Sigma -
i g^\prime \Sigma \frac{\tau^3}{2} B_\mu \; .
\end{equation}

We focused our attention on genuine photonic quartic interactions,
{\em i.e.}  the new interactions do not exhibit a triple gauge boson
vertex associated to them.  In our framework, genuine quartic
operators appear at next-to-leading order (${\cal O}(p^4)$), however,
there is no genuine photonic quartic interaction at this order.
Therefore, we considered the next order (${\cal O}(p^6)$).  There are
14 effective photonic operators which respect $SU(2)_c$ custodial
symmetry as well as ${\cal C}$ and ${\cal P}$,
\begin{eqnarray}
{\cal L}=&
\frac{\displaystyle g^2}{\displaystyle \Lambda^2}\Big[ &
k_0^w {\rm Tr} (\hat W_{\mu \nu} \hat W^{\mu \nu}) {\rm Tr}(V^\alpha V_\alpha)+ 
k_c^w {\rm Tr} (\hat W_{\mu \nu} \hat W^{\mu \alpha}) {\rm Tr}(V^\nu V_\alpha)+ 
k_1^w {\rm Tr} (\hat W_{\mu \nu}  V^\alpha) 
{\rm Tr} (\hat W^{\mu \nu} V_\alpha)\nonumber \\ 
&& 
+ k_2^w {\rm Tr} (\hat W_{\mu \nu} V^\nu) {\rm Tr} (\hat W^{\mu \alpha} V_\alpha) +
k_3^w {\rm Tr} (\hat W_{\mu \nu} V_\alpha) {\rm Tr} (\hat W^{\mu \alpha} V^\nu) 
\Big] + \nonumber\\
&\frac{\displaystyle{g'}^2}{\displaystyle\Lambda^2}\Big[& 
k_0^b {\rm Tr} (\hat B_{\mu \nu} \hat B^{\mu \nu}) {\rm Tr}(V^\alpha V_\alpha)+ 
k_c^b {\rm Tr} (\hat B_{\mu \nu} \hat B^{\mu \alpha}) {\rm Tr}(V^\nu V_\alpha) 
\nonumber \\
&& + k_1^b {\rm Tr} (\hat B_{\mu \nu} V^\alpha) 
{\rm Tr} (\hat B^{\mu \nu} V_\alpha)+ 
k_2^b {\rm Tr} (\hat B_{\mu \nu} V^\nu) {\rm Tr} (\hat B^{\mu \alpha} V_\alpha) 
\Big] + \label{lagrangian}  \\
&\frac{\displaystyle gg'}{\displaystyle\Lambda^2}\Big[ &
k_0^m {\rm Tr} (\hat W_{\mu \nu} \hat B^{\mu \nu}) {\rm Tr}(V^\alpha V_\alpha)+ 
k_c^m {\rm Tr} (\hat W_{\mu \nu} \hat B^{\mu \alpha}) {\rm Tr}(V^\nu V_\alpha)+ 
k_1^m {\rm Tr} (\hat W_{\mu \nu} V^\alpha) 
{\rm Tr} (\hat B^{\mu \nu} V_\alpha) \nonumber \\
&& 
+k_2^m {\rm Tr} (\hat W_{\mu \nu} V^\nu) {\rm Tr} (\hat B^{\mu \alpha} V_\alpha) +
k_3^m {\rm Tr} (\hat W_{\mu \nu} V_\alpha) {\rm Tr} (\hat B^{\mu \alpha} V^\nu) 
\Big] \; , \nonumber
\end{eqnarray}
where $V_\mu \equiv \left ( D_\mu\Sigma \right ) \Sigma^\dagger$,
$\hat{B}_{\mu \nu} = \tau^3 B_{\mu \nu}/2$, and $\hat{W}_{\mu \nu} =
\tau^a W^a_{\mu \nu}/2$, with $B_{\mu \nu}$ and $ W^a_{\mu \nu}$ being
respectively the $U(1)_Y$ and $SU(2)_L$ field strength tensors.  Here,
$e$ is the electromagnetic coupling, $g=e/\sin \theta_W=e/s_w$, and
$g' = g/c_w$ with $c_w = \sqrt{1-s_w^2}$.  $\Lambda$ is a mass scale
characterizing the NP.

It is interesting to express the effective interactions in 
(\ref{lagrangian}) in terms of independent Lorentz
structures. The lowest order effective
$W^+W^-\gamma\gamma$ and $ZZ\gamma\gamma$ interactions are described
in terms of four Lorentz invariant structures
\begin{eqnarray}
{\cal{W}}_0^\gamma &=& - \frac{e^2 g^2}{2}
F_{\mu\nu} F^{\mu\nu} W^{+ \alpha} W^-_\alpha \; ,
%\nonumber 
\label{lis_w0a} \\
{\cal{W}}_c^\gamma &=& - \frac{e^2 g^2}{4} 
F_{\mu\nu} F^{\mu\alpha} (W^{+ \nu} W^-_\alpha + W^{- \nu} W^+_\alpha)
\; ,
%\nonumber 
\label{lis_wca} \\
{\cal{Z}}_0^\gamma &=& - \frac{e^2 g^2}{4 c_w^2}
F_{\mu\nu} F^{\mu\nu} Z^\alpha Z_\alpha
\; , 
%\nonumber
\label{lis_z0a} \\
{\cal{Z}}_c^\gamma &=& - \frac{e^2 g^2}{4 c_w^2} 
F_{\mu\nu} F^{\mu\alpha} Z^\nu Z_\alpha \; ,
%\nonumber
\label{lis_zca}
\end{eqnarray}
while  the lowest order effective the $Z Z Z \gamma$ interactions are
given by
\begin{eqnarray}
{\cal{Z}}_0^Z &=& - \frac{e^2 g^2}{2 c_w^2 }
F_{\mu\nu} Z^{\mu\nu} Z^\alpha Z_\alpha  \; ,
%\nonumber
\label{lis_z0z} \\
{\cal{Z}}_c^Z &=& - \frac{e^2 g^2}{2 c_w^2 } 
F_{\mu\nu} Z^{\mu\alpha} Z^\nu Z_\alpha \; .
%\nonumber
\label{lis_zcz}
\end{eqnarray}
The remaining  $W^+ W^- Z \gamma$ interactions are parameterized as
\begin{eqnarray}
{\cal{W}}_0^Z &=& -  e^2 g^2  
F_{\mu\nu} Z^{\mu\nu} W^{+ \alpha} W^-_\alpha  \; , 
%\nonumber 
\label{lis_w0z} \\
{\cal{W}}_c^Z &=& - \frac{e^2 g^2}{2}
F_{\mu\nu} Z^{\mu\alpha} (W^{+ \nu} W^-_\alpha + W^{- \nu} W^+_\alpha)
\; , 
%\nonumber 
\label{lis_wcz} \\
{\cal{W}}_1^Z &=& - \frac{e^2 g^2}{2 c_w s_w}
F^{\mu\nu}(W^+_{\mu\nu} W^-_\alpha Z^\alpha +  
W^-_{\mu\nu} W^+_\alpha Z^\alpha)  \; , 
%\nonumber 
\label{lis_w1z} \\
{\cal{W}}_2^Z &=& - \frac{e^2 g^2}{2 c_w s_w }
F^{\mu\nu}(W^+_{\mu\alpha} W^{- \alpha} Z_\nu + 
W^-_{\mu\alpha} W^{+ \alpha} Z_\nu)  \; , 
%\nonumber 
\label{lis_w2z} \\
{\cal{W}}_3^Z &=& - \frac{e^2 g^2}{2 c_w s_w } 
F^{\mu\nu}(W^+_{\mu\alpha} W^-_\nu Z^\alpha + 
W^-_{\mu\alpha} W^+_\nu Z^\alpha) \; . 
%\nonumber 
\label{lis_w3z}
\end{eqnarray}
The Feynman rules for the quartic couplings induced by the above
operators can be found in Ref.~\cite{Belanger:1999aw}.

Eq.~(\ref{lagrangian}) can be conveniently rewritten in terms of the
above independent Lorentz structures, neglecting possible $4W$, $4Z$,
$WWZZ$ as well as Goldstone boson vertices, as
\begin{eqnarray}
{\cal L}=&&
\frac{k_0^\gamma}{\displaystyle\Lambda^2} 
\big({\cal Z}^\gamma_0+{\cal W}_0^\gamma\big)+
\frac{k_c^\gamma}{\displaystyle\Lambda^2} 
\big({\cal Z}^\gamma_c+{\cal W}_c^\gamma\big)+
\frac{k_1^\gamma}{\displaystyle\Lambda^2} 
{\cal Z}^\gamma_0 + \frac{k_{23}^\gamma}{\displaystyle\Lambda^2} 
{\cal Z}^\gamma_c
\label{group}  \\
&& 
+ 
\frac{k_0^Z}{\displaystyle\Lambda^2} {\cal Z}^Z_0 +
\frac{k_c^Z}{\displaystyle\Lambda^2} {\cal Z}^Z_c +
\sum_i\frac{k_i^W}{\displaystyle\Lambda^2} {\cal W}^Z_i \nonumber
\end{eqnarray}
with 
\begin{eqnarray}
&&k^\gamma_i=k_i^w+k_i^b+k_i^m \;\;{\rm for}\;\; i=0,c,1 \;, \label{jj1}\\
&&k^\gamma_{23}=k_2^w+k_2^b+k_2^m+ k_3^w+k_3^m   \;, 
\label{jj2}\\
&& k_0^Z=\frac{c_w}{s_w} (k_0^w+k_1^w)-\frac{s_w}{c_w}(k_0^b+k_1^b)
+c_{zw} (k_0^m+k_1^m)\; , \label{ll1} \\
&& k_c^Z=\frac{c_w}{s_w} (k_c^w+k_2^w+k_3^w)-\frac{s_w}{c_w}(k_c^b+k_2^b)
+c_{zw} (k_c^m+k_2^m+k_3^m)\; , \label{ll2} \\
&& k_0^W=\frac{c_w}{s_w} k_0^w-\frac{s_w}{c_w} k_0^b
+c_{zw} k_0^m \; , \label{ll3} \\
&& k_c^W=\frac{c_w}{s_w} k_c^w-\frac{s_w}{c_w} k_c^b
+c_{zw} k_c^m \; , \label{ll4} \\
&&k^W_i=k_i^w+\frac{1}{2}k_i^m \;\;{\rm for}\;\;  i=1,2,3 \, \label{ll5}
\end{eqnarray}
and $c_{zw}=(c_w^2-s_w^2)/(2c_ws_w)$.

Before we study the phenomenological consequences of anomalous quartic
vertices, we should stress that the effective Lagrangian
(\ref{group}) can also be obtained using a linear
representation of the $SU(2)_L \times U(1)_Y$ gauge symmetry with
the presence of a Higgs boson in the
spectrum~\cite{Belanger:1999aw}.  However, in this case, the lowest
order terms that can be written are of dimension 8 and they lead to
different relations between the couplings associated to the
independent Lorentz structures. Moreover, they generate 
both photonic and non-photonic genuine quartic vertices whose
strength is in general related, unlike in the non-linear case.

%---------------------------------------------------------------------
\section{Signals and backgrounds}

In this work we study the reactions (\ref{jj}) and (\ref{ll}) at the
LHC. We evaluated numerically the helicity amplitudes of all the SM
subprocesses leading to the $jj \gamma\gamma$ and $jj\gamma l^+ l^-$
final states where $j$ can be either a gluon, a quark or an anti-quark
in our partonic Monte Carlo.  The SM amplitudes were generated using
Madgraph~\cite{mad} in the framework of Helas~\cite{helas} routines.
The anomalous interactions arising from the Lagrangian
(\ref{lagrangian}) were implemented as subroutines and were included
accordingly. We consistently took into account the effect of all
interferences between the anomalous and the SM amplitudes and did not
use the narrow--width approximation for the vector boson propagators.
We considered a center--of--mass energy of 14 TeV and an integrated
luminosity of 100 fb$^{-1}$ for LHC.

It is important to note that the operators in Eq.~(\ref{lagrangian})
lead to tree--level unitarity violation in $2\to2$ processes at high
energies~\cite{our:quartic}.  The standard procedure to avoid this
unphysical behavior of the cross section and to obtain meaningful
limits is to multiply the anomalous couplings ($k_i^j$) by a form
factor
\begin{equation}
k_i^j \longrightarrow \left(1 +
\frac{m_{\gamma\gamma}^2}{\Lambda_u^2}\right)^{-n} \times k_i^j
\; ,
\label{ff_jj}
\end{equation}
where $m_{\gamma\gamma}$ is the invariant mass of the final state
photon pair in subprocesses like $ZZ \to \gamma \gamma$ and $WW \to
\gamma \gamma$. For subprocesses of the type $ZZ \to Z \gamma \to l^+
l^- \gamma$ and $WW \to Z \gamma \to l^+ l^- \gamma$ the anomalous
couplings are multiplied by a form factor
\begin{equation}
k_i^j \longrightarrow \left(1 +
\frac{m_{\ell^+ \ell^-\gamma}^2}{\Lambda_u^2}\right)^{-n} \times k_i^j
\; ,
\label{ff_ll}
\end{equation}
where $m_{\ell^+ \ell^- \gamma}$ is the invariant mass of the final
state lepton pair plus a photon. Of course using this procedure the
limits become dependent on the exponent $n$ and the scale $\Lambda_u$
which is not longer factorizable.  In our calculations, we
conservatively choose $n=5$ and $\Lambda_u$ = 2.5 TeV for the LHC.

At $e^+e^-$ colliders the center--of--mass energy is fixed and the
introduction of the form factors (\ref{ff_jj}) and (\ref{ff_ll}) is
basically equivalent to a rescaling of the anomalous couplings
$k_i^j$, therefore we should perform this rescaling when comparing
results obtained at hadron and $e^+e^-$ colliders. For example, the
LEP limits should be weakened by a factor $\simeq 1.6$ for our choice
of $n$ and $\Lambda_u$.

Altogether the cross sections for processes (\ref{jj}) and (\ref{ll}) 
can be written as 
\begin{equation}
\sigma \equiv \sigma_{\text{sm}} + \frac{k_i^j}{\Lambda^2}~ 
  \sigma_{\text{inter}} + \frac{{k_i^j}^2}{\Lambda^4}~ 
\sigma_{\text{ano}} \; ,  
\label{crosssection}
\end{equation}
where $\sigma_{\text{sm}}$, $\sigma_{\text{inter}}$, and
$\sigma_{\text{ano}}$ are, respectively, the SM cross section,
interference between the SM and the anomalous contribution and the
pure anomalous cross section.

%---------------------------------------------------------------------
\subsection{$p + p \to j + j + \gamma + \gamma$}

This process receives contributions from $ZZ\gamma\gamma$ and
$WW\gamma\gamma$ vertices which get modified by all operators in
Eq.~(\ref{lagrangian}). However, as seen in the first line in
Eq.~(\ref{group}) there are only four independent Lorentz invariant
structures contributing to this process which, consequently, is only
able to give information on the four linear combinations of anomalous
couplings corresponding to the four coefficients, $k^\gamma_i$,
($i=0,c,1,23$) defined in Eqs.~(\ref{jj1}) and~(\ref{jj2}).

Process (\ref{jj}) receives contributions from $W^*$ and $Z^*$
productions in association to photons as well as from $WW$ and $ZZ$
fusion processes
\begin{equation}
p + p  \to q + q + (W^* + W^* \mbox{ or } Z^* + Z^*) \to q + q + \gamma + 
\gamma \; .
\label{vbf}
\end{equation}     
In order to reduce the enormous QCD background we must exploit the
characteristics of the WBF reactions.  The main feature of WBF
processes is a pair of very far forward/backward tagging jets with
significant transverse momentum and large invariant mass between them.
Therefore, we required that the jets should comply with
\begin{eqnarray}
&&p_{T}^{j_{1(2)}} >  40~ (20)  \; \text{GeV} \;\;\;\; \hbox{,} \;\;\;\;
|\eta_{j_{(1,2)}}| <  5.0  \; ,
\nonumber \\
&& |\eta_{j_{1}} - \eta_{j_{2}}| >  4.4  \;\;\;\; \hbox{,} \;\;\;\;\;\;
\eta_{j_{1}} \cdot \eta_{j_{2}} < 0  \;\;\;\; \hbox{and} 
\label{cuts_jj1} \\
&& \Delta R_{jj} >  0.7\; .
 \nonumber 
\end{eqnarray}
Furthermore, the photons are central, typically being between the
tagging jets. So, we require that the photons satisfy
\begin{eqnarray}
&& E_{T}^{\gamma_{(1,2)}}  >   25  \; \text{GeV} 
 \;\;\;\; \hbox{,} \;\;\;\;
|\eta_{\gamma_{(1,2)}}|   <  2.5  \; ,
\nonumber 
\label{cuts_jj2}
\\
&& \text{min}\{\eta_{j_{1}}, \eta_{j_{2}} \}  + 0.7 <
 \eta_{\gamma_{(1,2)}} 
< \text{max}\{ \eta_{j_{1}}, \eta_{j_{2}} \} - 0.7 \; , 
 \\
&& \Delta R_{j\gamma} > 0.7  \;\;\;\; \hbox{and}
\;\;\;\; \Delta R_{\gamma \gamma} > 0.4
\; .
 \nonumber 
\end{eqnarray}

\begin{figure}
\protect
\centerline{\mbox{\psfig{file=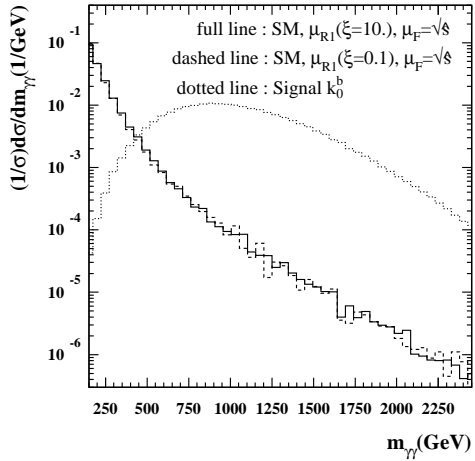,angle=0,width=0.5\textwidth}}}
\vskip -1. cm 
\caption{Normalized invariant mass distribution of the $\gamma\gamma$
pair for the reaction $p  p \to \gamma \gamma j j$.}
\label{im_gg_signal}
\end{figure}

Further reduction of the SM background can be achieved by a cut in the
invariant mass distribution of the $\gamma\gamma$ pairs. As
illustrated in Fig.~\ref{im_gg_signal}, the invariant mass
distribution for the SM background contribution is a decreasing
function of the $\gamma\gamma$ invariant mass while the anomalous
contribution first increases with the $\gamma\gamma$ invariant mass
reaching its maximum value at $m_{\gamma \gamma} \sim 1000$ GeV and
then decreases. Consequently, in order to enhance the WBF signal for
the anomalous couplings we imposed the following additional cut in the
diphoton invariant mass spectrum
\begin{eqnarray}
400 \text{ GeV } \leq m_{\gamma \gamma} \leq  2500
\text{ GeV.}
\label{cuts_jj3}
\end{eqnarray}

We present in Table \ref{tabano_aa} the values for $\sigma_{\text{ano}}$ for
each of the independent linear combination of anomalous couplings in
Eqs.~(\ref{jj1})~and~(\ref{jj2}) after applying the cuts in
Eqs.~(\ref{cuts_jj1})--(\ref{cuts_jj3}).  These results were obtained using
$\sqrt{\hat{s}}$ as the factorization scale in the parton distribution
functions.  We have further assumed a 85\% detection efficiency of isolated
photons, leptons and jet-tagging.  With this the efficiency for reconstructing
the final state $j + j + \gamma + \gamma$ is $(0.85)^4 \approx$ 52\% which is
included in the results presented in Tables~\ref{tabano_aa}
and~\ref{tabsm_aa}.  The interference terms ($\sigma_{\text{inter}}$) between
the anomalous and SM amplitudes turn out to be negligible.  As expected the
$WW$ fusion process due to $\wng$ ($\wcg$) leads to a larger anomalous
contribution (by a factor $\simeq 2.5$) than the $ZZ$ fusion ones due to the
$\zng$ ($\zcg$).

\begin{table}
\begin{tabular}{||c||c||}
\hline
\hline 
Coupling Constant & $\sigma_{\text{ano}}$ (pb $\times$ GeV$^4$)  \\ 
\hline
\hline 
$k^\gamma_0$ & $ 2.1 \times 10^{7} $ \\ 
\hline  
$k^\gamma_c$ & $ 1.5 \times 10^{6} $ \\ 
\hline  
$k^\gamma_1$ & $ 6.0 \times 10^{6} $ \\ 
\hline  
$k^\gamma_{23}$ & $ 4.3 \times 10^{5} $ \\
\hline
\hline 
\end{tabular}
\medskip
\caption{Results for $\sigma_{\text{ano}}$ for process 
Eq.~(\protect{\ref{jj}}) [see Eq.~(\protect{\ref{crosssection}})]. 
We considered $n = 5$ and $\Lambda_u=2.5$ TeV; see Eq.~(\protect{\ref{ff_jj}}).
All results include the effect of the cuts in 
Eq.~(\ref{cuts_jj1}), (\ref{cuts_jj2}), and (\ref{cuts_jj3}) as well as
 photon detection and jet-tagging efficiencies.} 
\label{tabano_aa}
\end{table}

The evaluation of the SM background ($\sigma_{\text{sm}}$) deserves
some special care since it has a large contribution from QCD
subprocesses whose size depends on the choice of the renormalization
scale used in the evaluation of the QCD coupling constant,
$\alpha_s(\mu_R)$, as well as on the factorization scale $\mu_F$ used
for the parton distribution functions.  To estimate the uncertainty
associated with these choices, we have computed $\sigma_{\text{sm}}$
for two sets of renormalization scales, which we label as
$\mu_{R1,2}(\xi)$, and for several values of $\mu_F$.  $\mu_{R1}(\xi)$
is defined such that $\alpha_s^2(\mu_{R1}(\xi))= \alpha_s(\xi
p_T^{j1})\alpha_s(\xi pt_T^{j2})$ where $p_T^{j1}$ and $p_T^{j2}$ are
the transverse momentum of the tagging jets and $\xi$ is a free
parameter varied between 0.1 and 10. The second choice of
renormalization scale set is $\mu_{R2} (\xi) =\xi \sqrt{\hat s}/2$,
with $\sqrt{\hat{s}}$ being the subprocess center--of-mass energy.

\begin{table}
\begin{tabular}{||c||c|c|c||c|c|c||}
\hline 
\hline 
&  \multicolumn{6}{c||}{$\sigma_{\text{sm}}$ (fb)} \\ 
\hline
& \multicolumn{3}{c||}{$\mu_R=\mu_{R1}(\xi)$ } 
& \multicolumn{3}{c||}{$\mu_R=\mu_{R2}(\xi)$ } \\ 
\hline
$\xi$ & 
$\mu_F=\sqrt{\hat s} $& $\mu_F=p^T_{\rm min}$ 
& $\mu_F=\sqrt{\hat s}/10 $ & 
$\mu_F=\sqrt{\hat s} $& $\mu_F=p^T_{\rm min}$ 
& $\mu_F=\sqrt{\hat s}/10 $\\
\hline
\hline 
0.10 & 3.2 & 5.3 & 4.1 & 1.3 & 2.2 & 1.7\\
\hline 
0.25 & 2.2 & 3.6 & 2.8 & 1.1 & 1.9 & 1.4\\ 
\hline 
1.00 & 1.4 & 2.4 & 1.9 & 0.91 & 1.5 & 1.2\\ 
\hline 
4.00 & 1.1 & 1.8 & 1.4 & 0.78 & 1.3 & 1.0\\ 
\hline 
10.0 & 0.94 & 1.6 & 1.2 & 0.71 & 1.2 & 0.96
\\ 
\hline 
\hline 
\end{tabular}
\medskip
\caption{Results for $\sigma_{\text{sm}}$ for process 
Eq.~(\protect{\ref{jj}}); see
Eq.~(\protect{\ref{crosssection}}) and text for details.
All results include the effect of the cuts in 
Eq.~(\ref{cuts_jj1}), (\ref{cuts_jj2}) and (\ref{cuts_jj3}) as well as
photon detection and jet-tagging efficiencies.}
\label{tabsm_aa}
\end{table}

In Table \ref{tabsm_aa} we list $\sigma_{\text{sm}}$ for the two sets
of renormalization scales %with different values of the factor $\xi$
and for three values of the factorization scale $\mu_F=\sqrt{\hat s}$,
$\sqrt{\hat s}/10$, and $p^T_{\rm min}$ where $p^T_{\rm min}={\rm min}
(p_T^{j1},p_T^{j2})$. As shown in this table, we find that the
predicted SM background can change by a factor of $\sim 8$ depending
on the choice of the QCD scales.  These results indicate that to
obtain meaningful information about the presence of anomalous
couplings one cannot rely on the theoretical evaluation of the
background.  Instead one should attempt to extract the value of the SM
background from data in a region of phase space where no signal is
expected and then extrapolate to the signal region.

In looking for the optimum region of phase space to perform this
extrapolation, one must search for kinematic distributions for which
(i) the shape of the distribution is as independent as possible of the
choice of QCD parameters. Furthermore, since the electroweak and QCD
contributions to the SM backgrounds are of the same order~\cite{info},
this requires that (ii) the shape of both electroweak and QCD
contributions are similar. Several kinematic distributions verify
condition (i), for example, the azimuthal angle separation of the two
tagging jets which was proposed in Ref.~\cite{Eboli:2000ze} to reduce
the perturbative QCD uncertainties of the SM background estimation for
invisible Higgs searches at LHC.  However, the totally different shape
of the electroweak background in the present case, renders this
distribution useless.

\begin{figure}
\protect
\centerline{\mbox{\psfig{file=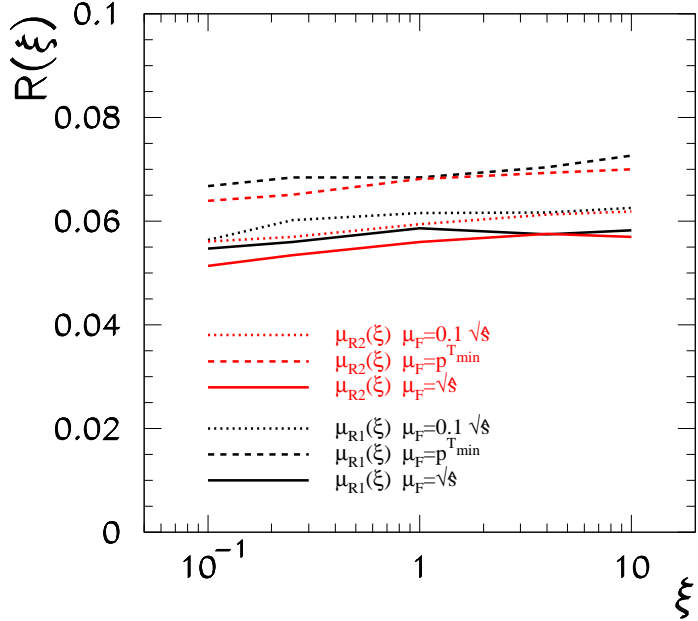,angle=0,width=0.6\textwidth}}}
\caption{Ratio $R(\xi)$ defined in Eq.~(\ref{rxijj})  
for the process $p p \to \gamma  \gamma  j  j$ at LHC.}
\label{ratio}
\end{figure}

We found that the best sensitivity is obtained by using the
$\gamma\gamma$ invariant mass.  As can be seen in
Fig.~\ref{im_gg_signal}, the shape of the SM distribution is quite
independent of the choice of the QCD parameters.  As a consequence
most of the QCD uncertainties cancel out in the ratio
\begin{equation}
R(\xi)=\frac{\sigma({400\text{ GeV }<m_{\gamma \gamma}<2500\text{ GeV }}) }
{\sigma({100\text{ GeV }<m_{\gamma \gamma}<400\text{ GeV }}) }\;\;.
\label{rxijj}
\end{equation}
This fact is illustrated in Fig.~\ref{ratio} where we plot the value
of the ratio $R(\xi)$ for different values of the renormalization and
factorization scales.  The ratio $R$ is almost invariant under changes
of the renormalization scale, showing a maximum variation of the order
of $\pm6$\% for a fixed value of the factorization scale. On the other
hand, the uncertainty on the factorization scale leads to a maximum
variation of 12\% in the background estimation.  We have also verified
that different choices for the structure functions do not affect these
results.

Thus the strategy here proposed is simple: the experiments should
measure the number of events in the $\gamma\gamma$ invariant mass
window $100 < m_{\gamma\gamma}< 400$ GeV and extrapolate the results
for higher invariant masses using perturbative QCD.  According to the
results described above we can conservatively assign a maximum ``QCD''
uncertainty ($ {\rm QCD}_{\rm unc}$) of $\pm$ 15\% to this
extrapolation.

In order to estimate the attainable sensitivity to the anomalous
couplings we assume that the observed number of events is compatible
with the expectations for $\mu_{R1}(\xi=1)$ and $\mu_F=\sqrt{\hat s}$,
so the observed number of events in the signal region coincides with
the estimated number of background events obtained from the
extrapolation of the observed number of events in the region where no
signal is expected; for this choice the number of expected background
events is $N_{\rm back}=\sigma_{\text{sm}} {\cal L}$ where ${\cal L}$
stands for the integrated luminosity.  For an integrated luminosity of
100 fb$^{-1}$ for LHC, this corresponds to $N_{\rm back}=143$.
Moreover, we have added in quadrature the statistical error and the
QCD uncertainty associated with the backgrounds. Therefore, the 95\%
limits on the quartic couplings can be obtained from the condition
\begin{equation}
N_{\rm ano}\,=\, \frac{k^2_j}{\Lambda^4} \,\times {\cal L } \,\times \,
\sigma_{\rm ano} 
\leq  1.95 \sqrt{N_{\rm back}+ (N_{\rm back}\times {\rm QCD}_{\rm
    unc})^2} \; .
\end{equation}

For the sake of completeness we show the results on the expected
sensitivity using purely statistical errors and for two values of
${\rm QCD}_{\rm unc}$: our most conservative estimate [15 \%], and a
possible reduced uncertainty (7.5 \%), which could be attainable provided
NLO QCD calculations are available. Assuming
that only one operator is different from zero, so no cancellations 
are possible, we find
\begin{eqnarray}
|k_0^{w,b,m}/\Lambda^2| < 3.3\, (3.9)\,[4.8] \times 10^{-6} \text{ GeV}^{-2} \;,
\nonumber \\  
|k_c^{w,b,m}/\Lambda^2| < 1.3\,(1.5)\,[1.8] \times 10^{-5} \text{ GeV}^{-2} \;,
\nonumber \\  
|k_1^{w,b,m}/\Lambda^2| < 6.2\,(7.2)\,[8.9] \times 10^{-6} \text{ GeV}^{-2} \;,
\label{aalimits} \\  
|k_2^{w,b,m}/\Lambda^2 | < 2.3\,(2.7)\,[3.3] \times 10^{-5} \text{ GeV}^{-2} \;,
\nonumber \\
|k_3^{w,m}/\Lambda^2| < 2.3\,(2.7)\,[3.3] \times 10^{-5} \text{ GeV}^{-2} \;.
\nonumber
\end{eqnarray}
We notice that the constraint on $k_2^{w,b,m}$ and  $k_3^{w,m}$
are exactly the same as they both modify in the same form and amount
the process (\ref{jj}) as seen in Eq.~(\ref{group}).  

Finally, let us comment that the limits on $k_0^{w,b,m}/\Lambda^2$ and
$k_c^{w,b,m}/\Lambda^2$ can be directly translated on constraints on
the coefficients $a_{0,c}$, of the operators introduced in
Ref.~\cite{bel:bou} with the substitution $a_{0,c}=4 g^2
k_{0,c}^\gamma$ (see Eq.~(\ref{jj1})).

%---------------------------------------------------------------------
\subsection{$p + p \to j + j + \gamma + l^+ + l^-$}

This process receives contribution from the four-gauge coupling
vertices $ZZZ\gamma$ and $WWZ\gamma$ as well as from $ZZ\gamma\gamma$
and $WW\gamma\gamma$.  We have imposed a minimal set of cuts to
guarantee that the photons, charged leptons and jets are detected and
isolated from each other:
\begin{eqnarray}
&& p_{T}^{j_{1(2)}} \geq  40(20) \text{ GeV } \;\;\;\; \hbox{,}\;\;\;\;
p_{T}^{\ell} \geq  25 \text{ GeV } \;\;\;\; \hbox{,}\;\;\;\;
E_{T}^{\gamma} \geq  25 \text{ GeV }
\; , \nonumber \\
&& |\eta_{\gamma,\ell}| \leq 2.5  \;\;\;\; \hbox{,}\;\;\;\; 
|\eta_{j_{(1,2)}}|     < 5.0  
\; ,\nonumber 
 \\
&& |\eta_{j_{1}} - \eta_{j_{2}}| > 4.4  \;\;\;\; \hbox{,}\;\;\;\; 
\eta_{j_{1}} . \eta_{j_{2}} < 0 
\; , 
\label{cuts_ll} 
\\
&& \text{min}\{\eta_{j_{1}}, \eta_{j_{2}} \}  + 0.7 <
\eta_{\gamma,\ell} 
< \text{max}\{ \eta_{j_{1}}, \eta_{j_{2}} \} - 0.7 
\; , \nonumber 
\\
&& \Delta R_{jj~(j\gamma,j\ell)} > 0.7
  \;\;\;\; \hbox{,}\;\;\;\;
\Delta R_{\ell^+\ell^-(\gamma\ell)} > 0.4
\; . \nonumber 
\end{eqnarray}
Furthermore, in order to single out the events containing $Z^0$ bosons
and to enhance the WBF signal for the anomalous couplings $ZZZ\gamma$
and $WWZ\gamma$ we have imposed the following additional cuts on the
lepton-lepton ($m_{\ell \ell}$) and lepton-lepton-photon ($m_{\gamma
  \ell \ell}$) invariant masses:
\begin{equation}
|m_{\ell \ell} - M_Z| \leq  20 \text{ GeV }
 \;\;\;\; \hbox{and}\;\;\;\; 
400 \text{ GeV } \leq  m_{\gamma \ell \ell}  \leq  2500 \text{ GeV.}
\label{cuts_ll2}
\end{equation}

In Table \ref{tabano_eea} we display the values of
$\sigma_{\text{ano}}$ after cuts for each anomalous couplings $k_i^j$
in Eq.~(\ref{lagrangian}), with $\mu_F = \sqrt{\hat{s}}$.  
These results include the effect of
detection and tagging efficiencies; 85\% efficiency for detecting
isolated photons and leptons and for tagging a jets.  With this, the
efficiency for reconstructing the final state $j + j + \gamma + l^+\,
l^-$ is $(0.85)^5 \approx$ 44\%.  We have added the contributions from
final states containing electrons and muons. Once again, we verified
that the interference terms $\sigma_{\text{inter}}$ are negligible.

A detailed study of the results in terms of the different Lorentz structures
involved show that the invariant mass cut on the lepton-lepton invariant mass
suppresses the contributions from the $W^+ W^- \gamma \gamma$ Lorentz
structures ${\cal W}_0^\gamma$ and ${\cal W}_c^\gamma$ in relation to those
containing the $VVZ\gamma$ and $ZZ\gamma\gamma$ quartic vertices ($V=W$ or
$Z$). However, we find that none of the Lorentz structures involving these
vertices is clearly dominant and that there are important interference effects
between the different Lorentz structures contributing to the same anomalous
operator, which are the order of 10--30\% and can be destructive or
constructive.

\begin{table}
\begin{tabular}{||c||c||}
\hline
\hline
Coupling Constant & $\sigma_{\text{ano}}$ (pb $\times$ GeV$^4$)  \\
\hline
\hline 
$k_0^w$ & $ 4.6 \times 10^{7} $ \\ 
\hline  
$k_c^w$ & $ 9.2 \times 10^{6} $ \\ 
\hline  
$k_1^w$ & $ 2.9 \times 10^{7} $ \\ 
\hline  
$k_2^w$ & $ 1.3 \times 10^{7} $ \\ 
\hline  
$k_3^w$ & $ 1.0 \times 10^{7} $ \\ 
\hline   
\hline  
$k_0^b$ & $ 6.9 \times 10^{6} $ \\ 
\hline  
$k_c^b$ & $ 1.9 \times 10^{6} $ \\ 
\hline  
$k_1^b$ & $ 4.7 \times 10^{6} $ \\ 
\hline  
$k_2^b$ & $ 1.8 \times 10^{6} $ \\ 
\hline  
\hline  
$k_0^m$ & $ 1.1 \times 10^{7} $ \\ 
\hline  
$k_c^m$ & $ 3.2 \times 10^{6} $ \\ 
\hline  
$k_1^m$ & $ 9.0 \times 10^{6} $ \\ 
\hline  
$k_2^m$ & $ 4.3 \times 10^{6} $ \\ 
\hline 
$k_3^m$ & $ 3.6 \times 10^{6} $\\
\hline
\hline
\end{tabular}
\medskip
\caption{Results for $\sigma_{\text{ano}}$ for the process  
(\protect{\ref{ll}}); see Eq.~(\protect{\ref{crosssection}}). 
$\sigma_{\text{ano}}$ is obtained for the anomalous coupling
$k_i^j/\Lambda^2$ in units of GeV$^{-2}$. We considered $n = 5$ and
$\Lambda_u=2.5$ TeV; see Eq.~(\protect{\ref{ff_ll}}).}
\label{tabano_eea}
\end{table}

The evaluation of the SM background in this case is also subject
to QCD uncertainties, as in the previous reaction.  We found that
for our reference value $\mu_{R1}(\xi=1)$ and $\mu_F=\sqrt{\hat s}$
\begin{equation}
            \sigma_{\rm sm}= 0.10 \; {\rm fb.}
\label{smjjgll}
\end{equation}

Changes in the factorization and renormalization scales, can modify
this prediction by a factor $\sim$ 5. Thus, again, the best strategy
to accurately determine the sensitivity to the anomalous coupling is
to extract the value of the SM background from data in a region of
phase space where no signal is expected and then extrapolate to the
signal region. Following the discussion in the previous section, we
find that the $\ell^+\ell^-\gamma$ invariant mass distribution is
suitable to estimate the SM background and reduce the QCD
uncertainties.  We have defined the ratio
\begin{equation}
R(\xi)=\frac{\sigma({400\text{ GeV }<m_{\ell \ell \gamma}<2500\text{ GeV }}) }
{\sigma({100\text{ GeV }<m_{\ell \ell \gamma}<400\text{ GeV }}) }\;\;,
\end{equation}
and evaluated the behavior of $R(\xi)$ under changes of the renormalization
and factorization scales. We determined that $R(\xi)$ can be known within an
accuracy of $\pm$ 15\% when we use leading order calculations.

In order to extract the attainable limits on the anomalous couplings
we assumed a luminosity of ${\cal L}=100$ fb$^{-1}$ and that the
observed number of events is compatible with the expectations for
$\mu_{R1}(\xi=1)$ and $\mu_F=\sqrt{\hat s}$, {\em i.e.} the expected
number of background events in the signal region is $N_{\rm
  back}=10$. We have added to the statistical error associated with
this background the theoretical error associated to the uncertainty in
the extrapolation of the background.  However, given the limited
statistics, the sensitivity is dominated by the statistical error.
The 95\% C.L. constraints on the anomalous couplings are
\begin{eqnarray}
|k_0^w/\Lambda^2| < 1.2  \times 10^{-6} \text{ GeV}^{-2} 
\;\; ,  \nonumber \\
|k_c^w/\Lambda^2| < 2.8  \times 10^{-6} \text{ GeV}^{-2}
\;\; ,  \nonumber \\  
|k_1^w/\Lambda^2| < 1.5  \times 10^{-6} \text{ GeV}^{-2}
\;\; ,  \nonumber \\  
|k_2^w/\Lambda^2| < 2.3  \times 10^{-6} \text{ GeV}^{-2} 
\;\; ,  \nonumber \\  
|k_3^w/\Lambda^2| < 2.6 \times 10^{-6} \text{ GeV}^{-2} 
\;\; ,  \nonumber \\   
|k_0^b/\Lambda^2| < 3.2 \times 10^{-6} \text{ GeV}^{-2}
\;\; ,   \label{llalimits} \\  
|k_c^b/\Lambda^2| < 6.0  \times 10^{-6} \text{ GeV}^{-2} 
\;\; ,  \nonumber \\  
|k_1^b/\Lambda^2| < 3.8 \times 10^{-6} \text{ GeV}^{-2}
\;\; ,  \nonumber \\  
|k_2^b/\Lambda^2| < 6.3 \times 10^{-6} \text{ GeV}^{-2}
\;\; ,  \nonumber \\  
|k_0^m/\Lambda^2| < 2.6  \times 10^{-6} \text{ GeV}^{-2}
\;\; ,  \nonumber \\  
|k_c^m/\Lambda^2| < 4.7 \times 10^{-6} \text{ GeV}^{-2} 
\;\; ,  \nonumber \\  
|k_1^m/\Lambda^2| < 2.8  \times 10^{-6} \text{ GeV}^{-2}
\;\; ,  \nonumber \\  
|k_2^m/\Lambda^2| < 4.0 \times 10^{-6} \text{ GeV}^{-2} 
\;\; ,  \nonumber \\ 
|k_3^m/\Lambda^2| < 4.4 \times 10^{-6} \text{ GeV}^{-2} 
\;\; .  \nonumber
\end{eqnarray}
which have been obtained including a 15\% QCD uncertainty, However, to
the precision quoted, the impact of this uncertainty is minimal.

Comparing the limits in Eqs.~(\ref{llalimits}) with the corresponding ones
from the process (\ref{jj}) in Eq.~(\ref{aalimits}) we see that, despite the
limited statistics, the presence of the $VVZ\gamma$ vertex ($V=W$ or $Z$)
makes the process $p p \to j j \gamma l^+ l^-$ most sensitive to the presence
of NP leading to anomalous four-vector couplings which respect the $SU(2)_L
\times U(1)_Y$ gauge invariance as well as the $SU(2)_{\rm c}$ custodial
symmetry. One of the reasons for the process $p p \to j j \gamma l^+ l^-$ to
be more sensitive to anomalous interactions is that almost all Lorentz
structures lead similar contributions and that more Lorentz structures
contribute to this reaction than in $p p \to j j \gamma \gamma$ for a given
effective operator.

One must keep in mind, however, that the results in Eqs.~(\ref{aalimits}) and
(\ref{llalimits}) were obtained under the assumption that only one operator is
different from zero, so no cancellations were possible. If cancellations are
allowed the process (\ref{jj}) may become the most sensitive one to the
presence of the relevant photonic quartic operators.

%---------------------------------------------------------------------
\section{Discussion}

We are just beginning to test the SM predictions for the quartic vector boson
interactions. Due to the limited available center--of--mass energy, the first
couplings to be studied should contain photons. In particular, the direct
searches at LEPII have lead to constraints of the order
$|\frac{k_i^j}{\Lambda^2}|\lesssim {\cal O} (10^{-2}\; {\rm GeV^{-2})}$ for
the couplings in Eq.~(\ref{lagrangian}), and no significantly better
sensitivity is expected from searches at Tevatron.  Anomalous quartic
couplings contribute at the one--loop level to the $Z$ physics~\cite{our:vvv}
via oblique corrections as they modify the $W$, $Z$, and photon two--point
functions.  Consequently they can be indirectly constrained by precision
electroweak data to $|\frac{k_i^j}{\Lambda^2}|\lesssim {\cal O} (10^{-4}\;
{\rm GeV^{-2})}$.

Higher energy colliders will be able to test quartic gauge couplings
involving photons as well as to probe non-photonic vertices $VVV'V'$
($V,V'=W$ or $Z$)~\cite{vvvv}.  Even at LHC energies, due to phase
space limitations, the best experimental sensitivity is expected for
couplings involving photons which can be part of the final state.
Moreover, in the event that a departure from the SM predictions is
observed, inference of the underlying dynamics can only be obtained by
comparing the observations in different channels, for instance between
those involving triple and quartic-gauge couplings.  In this respect
it will also be important to know whether NP reveals itself in the
form of anomalous four-gauge couplings involving only weak gauge
bosons or in those involving photons or in both.  For instance, in the
framework of chiral lagrangians, where no light Higgs state is
observed, the photonic four vertices are expected to be suppressed
with respect to the non-photonic ones since they appear one order
higher in the momentum expansion.  An anomalous signal only in the
photonic couplings could indicate that there are additional symmetries
forbidding the non-photonic vertices.

With this motivation, in this work we have analyzed the production of
two jets in association with a photon pair, or with a photon and a
$\ell^+\ell^-$ pair at LHC as tests of anomalous bosonic quartic
couplings involving one or two photons.  In this study we have taken
careful account of the theoretical uncertainties associated with the
evaluation of the SM background.  We have proposed the best strategy
to estimate the expected SM background by extrapolation of the data
taken in a region of phase space where no signal is expected,
minimizing the theoretical uncertainty associated to this
extrapolation.  The final sensitivity to the different couplings is
given in Eqs.~(\ref{aalimits}) and (\ref{llalimits}).  In particular,
we found that in the framework of $SU(2)_L \times U(1)_Y$ gauge
invariant NP in which the deviations from the SM prediction for the
$VV\gamma\gamma$ vertices are related to the strength of the anomalous
$VVZ\gamma$ vertex, the process $p p\to j j \gamma l^+ l^-$ is the
most sensitive to all possible operators, despite the limited
statistics, barring possible cancellations. It can lead to
constraints $|\frac{k_i^j}{\Lambda^2}|\lesssim 1.2$--$6.3 \times
10^{-6}\; {\rm GeV^{-2}}$.

In conclusion we have shown that the study of the processes (\ref{jj})
and (\ref{ll}) at LHC can test quartic anomalous couplings that are
four orders of magnitude weaker than the existing limits from
direct searches and two orders of magnitude weaker than any
indirect constraints.  It is interesting to notice that if no signal
is found the LHC will lead to limits that are similar to the ones that
could be attainable at an $e^+e^-$ collider operating at $\sqrt{s} =
500 $ GeV with a luminosity of 300 fb$^{-1}$~\cite{Belanger:1999aw,nlc}.

%%%%%%%%%%%%%%%%%%%%%%%%%%%%%%%%%%%%%%%%%%%%%%%%%%%%%%%%%%%%%%%%%%%%%%

\acknowledgments

This work was supported by Conselho Nacional de Desenvolvimento
Cient\'{\i}fico e Tecnol\'ogico (CNPq), by Funda\c{c}\~ao de Amparo
\`a Pesquisa do Estado de S\~ao Paulo (FAPESP), by Programa de Apoio a
N\'ucleos de Excel\^encia (PRONEX). MCG-G acknowledges support from
National Science Foundation grant PHY0098527 and Spanish Grants No
FPA-2001-3031 and CTIDIB/2002/24.

%%%%%%%%%%%%%%%%%%%%%%%%%%%%%%%%%%%%%%%%%%%%%%%%%%%%%%%%%%%%%%%%%%%%%%


\begin{references}
\bibitem{info} The electroweak contribution to the total SM background
  is approximately 25\% for $\mu_{R1}(\xi=1)$ and $\mu_F =
  \sqrt{\hat{s}}$.
\end{references}
\end{document}